\documentclass[twocolumn,showpacs,preprintnumbers,amsmath,amssymb,superscriptaddress]{revtex4}

\usepackage{graphicx}
\usepackage{dcolumn}
\usepackage{bm}
\usepackage{hyperref}
\usepackage{latexsym}

\begin{document}

\title{Dynamical properties of model communication networks}
\author{Roger Guimer\`a}

\affiliation{Departament d'Enginyeria Qu\'{\i}mica, Universitat Rovira
i Virgili, 43007 Tarragona , Spain}

\author{Alex Arenas} 

\affiliation{Departament d'Enginyeria Inform\`atica i Matem\`atiques,
Universitat Rovira i Virgili, 43007 Tarragona, Spain}

\author{Albert D\'{\i}az-Guilera}

\affiliation{Departament d'Enginyeria Qu\'{\i}mica, Universitat Rovira
i Virgili, 43007 Tarragona , Spain}

\affiliation{Departament de F\'{\i}sica Fonamental, Universitat de
Barcelona, 08028 Barcelona, Spain}

\author{Francesc Giralt}

\affiliation{Departament d'Enginyeria Qu\'{\i}mica, Universitat Rovira
i Virgili, 43007 Tarragona , Spain}

\begin{abstract}
We study the dynamical properties of a collection of models for
communication processes, characterized by a single parameter $\xi$
representing the relation between information load of the nodes and
its ability to deliver this information. The critical transition to
congestion reported so far occurs only for $\xi=1$. This case is well
analyzed for different network topologies. We focus of the properties
of the order parameter, the susceptibility and the time correlations
when approaching the critical point. For $\xi<1$ no transition to
congestion is observed but it remains a cross-over from a low-density
to a high-density state. For $\xi>1$ the transition to congestion is
discontinuous and congestion nuclei arise.
\end{abstract}

\pacs{89.20.Ff,05.70.Jk,64.60.-i}

\date{\today}

\maketitle

\section{Introduction}

The interaction between the {\it elements} of social, technological,
biological, chemical and physical systems usually defines complex
networks. The study of topological properties in such networks has
recently generated a lot of interest among the scientific community
\cite{watts98,barabasi99,amaral00,albert02,dorogovtsev??}. Part of
this interest comes from the attempt to understand the behavior of
technology based communication networks such as the Internet
\cite{faloutsos99}, the World Wide Web \cite{albert99,huberman99},
e-mail networks \cite{ebel??} or phone call networks
\cite{adamic01}. The study of communication processes is also of
interest in other fields, notably in the design of organizations
\cite{radner93,decanio98}. It is estimated that more than one-half of
the U.S. work force is dedicated to information processing, rather
than to make or sell things in the narrow sense \cite{radner93}.

Tools taken from statistical mechanics are used to understand not only
the topological properties of these communication networks, but also
their dynamical properties. Particularly interesting is the phenomenon
of congestion. In has been observed, both in real networks
\cite{jacobson88} and in model communication networks,
\cite{ohira98,tretyakov98,arenas01,sole01} that the system only
behaves efficiently when the amount of information handled is small
enough. The network collapses above a certain threshold and some
information is accumulated and remains undelivered over large time
periods---or it is simply lost. This transition from a {\it free} to a
{\it congested} regime is indeed a phase transition and could be
related to the appearance of the $1/f$ noise observed in Internet flow
data \cite{csabai94,takayasu96}.

Understanding congestion is also important from an engineering point
of view. For instance, in October of 1986, during the first Internet
collapse reported in the literature, the speed of the connection
between the Lawrence Berkeley Laboratory and the UC at
Berkeley---separated by 200 meters--- dropped by a factor of 100
\cite{jacobson88}. Although the problem of congestion, and in
particular its prevention and control \cite{jacobson88}, has been
studied because of its implications in digital communications, a deep
understanding of the physics of congestion for general communication
processes and beyond particular protocols is still lacking.

A general collection of models that captures the essential features of
communication processes has been recently proposed \cite{arenas01}. In
these models, agents---nodes---are organized in a hierarchical
network---a Cayley tree--- and interchange information {\it
packets}. Each agent has a certain capability that decreases as the
number of packets to deliver increases. When the capability is
inversely proportional to the number of accumulated packets, a
continuous phase transition was found between the free and the
congested phases. This transition was characterized by means of an
order parameter.

The aim of this paper is to study the collection of models mentioned
above, since only hierarchical lattices and a very particular
congestion behavior---capability inversely proportional to the number
of accumulated packets---were considered. An extension to consider the
existence of cost to maintain communication channels has already been
done \cite{guimera01}. First, the network model is extended beyond
purely hierarchical Cayley trees. In particular, 1-dimensional and
2-dimensional lattices are considered. It should be noted that simple
models such as Cayley trees and regular lattices can capture the main
characteristics of the dynamics of information flow in complex
networks. Hierarchical trees have been considered to model the TCP/IP
protocol in the Internet \cite{csabai94,takayasu96}. On the other
hand, computer based communication networks have been described by
placing routers and servers in square lattices \cite{ohira98,sole01}.

Moreover it is shown that, independently of the topology, the
collection of models can be split into three groups according to how
the network collapses. In the first group, agents deliver more packets
as they are more congested---although their capability, as defined in
\cite{arenas01}, decreases---, and the network never collapses. In the
second group, agents deliver always the same number of packets
independently of their load---number of packets to deliver---; this
behavior leads to a continuous phase transition as reported for
hierarchical lattices in \cite{arenas01}. Finally, when agents deliver
fewer packets as their loads increase, the transition to the congested
phase is discontinuous and the network collapses in an inhomogeneous
way giving rise to congestion nuclei. To characterize these behaviors,
the order parameter defined in \cite{arenas01} is used, but also the
power spectrum of the fluctuations and a susceptibility-like
function. Most of the current effort is focused on the quantitative
study of the continuous transition and the critical behavior
associated to such a transition.

The paper is organized as follows. Section 2 presents the collection
of models and describes some details concerning the extension beyond
hierarchical lattices, including 1D and 2D lattices. In section 3,
these models are studied numerically and analytically. The scaling of
the critical point with the size of the system, the behavior of the
order parameter, and the divergence of the characteristic time at the
critical point are studied in detail. Finally, section 4 includes the
discussion of the results and the conclusions.

\section{The model}
The model considers three basic components: (i) the physical support
for the communication process---agents and communication channels---,
(ii) the discrete information packets that are interchanged, and (iii)
the limited capability of the agents to handle such packets.

\begin{figure}[t]
\includegraphics*[width=.95\columnwidth]{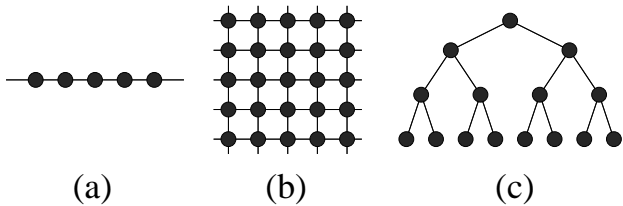}
\vspace*{-0.3cm}
\caption{Network topologies studied. (a) 1D lattice with $L=5$; (b) 2D
lattice with $L=5$ and $S=25$; and (c) $(2,3)$ Cayley tree with
branching $z=2$ and $m=3$ levels or generations.}
\label{nets}
\end{figure}

The communication network is mapped onto a graph where nodes mimic the
communicating agents (for instance, employees in a company, routers
and servers in a computer network, etc.) and the links between them
represent communication lines. In particular, the three different
topologies depicted in Fig.~\ref{nets} are considered: 1D lattices of
length $L$, 2D square lattices of side length $L$ and size $S=L^2$,
and hierarchical Cayley trees with branching $z$ and a number of
generations or levels $m$, hereafter denoted $(z,m)$.

The dynamics of the model is as follows. At each time step $t$, an
information packet is created at every node with probability
$p$. Therefore $p$ is the control parameter: small values of $p$
correspond to low density of packets and high values of $p$ correspond
to high density of packets. When a new packet is created, a
destination node, different from the origin node, is chosen randomly
in the network. Thus, during the following time steps
$t+1,\,t+2,\ldots ,\,t+T$, the packet travels toward its
destination. Once the packet reaches the destination node, it is
delivered and disappears from the network.

The time that a packet remains in the network is related not only to
the distance between the source and the target nodes, but also to the
amount of packets in its path. In particular, at each time step, all
the packets move from their current position, $i$, to the next node in
their path, $j$, with a probability $q_{ij}$. This probability
$q_{ij}$ is called the {\it quality of the channel} between $i$ and
$j$, and was defined in \cite{arenas01} as
\begin{equation}
q_{ij}=\sqrt{k_{i}k_{j}}\,,
\end{equation}
where $k_{i}$ represents the {\it capability} of agent $i$ at each
time step. The quality of a channel is, thus, the geometric average of
the capabilities of the two nodes involved, so that when one of the
agents has capability 0, the channel is disabled. High qualities
($q_{ij}\approx 1$) imply that packets move easily while low qualities
($q_{ij}\approx 0$) imply that it takes a long time for a packet to
jump from one node to the next. The general equation proposed for
$k_{i}$ is:
\begin{equation}
k_{i}=f(n_i)
\end{equation}
where $n_i$ is the total number of packets currently at node $i$. The
function $f(n)$ determines how the capability evolves when the number
of packets at a given node changes,
\begin{equation}
f(n)=\left\{ \begin{array}{lcl}
        1		&       \quad\mbox{for} &       n=0\\
        n^{-\xi}	&       \quad\mbox{for} &       n=1,2,3,\ldots
        \end{array}\right.
\label{f_1}
\end{equation}
with $\xi\geq0$. Equation (\ref{f_1}) defines a complete collection of
models with agents that behave differently depending on the exponent
$\xi$.

The average number of packets delivered during one time step by a node
$i$ to another node $j$ is proportional to
$n_i/(n_i^{\xi/2}n_j^{\xi/2})$. Assuming that $n_i\propto n_j$ the
former expression reads $n_i^{1-\xi}$. The proportionality is exact
both in the hierarchical lattice and in large enough 1D and 2D
lattices, where adjacent nodes are statistically equivalent.

Therefore, for $\xi<1$ the number of delivered packets increases with
the number of accumulated packets. For $\xi>1$ the number of delivered
packets decreases as the number of accumulated packets
increases. Finally, for the particular case $\xi=1$, the number of
delivered packets is independent of the number of accumulated
packets. Note that this particular case is consistent with simple
models of queues \cite{ohira98}.

The last point that needs to be explained to completely define the
current model is the routing algorithm or, in other words, the set of
rules that the nodes follow to select where to send a certain
packet. In all the topologies considered (1D, 2D and Cayley) the
packets follow paths of minimum length from their origin to their
destination---open boundary conditions are set in both 1D and 2D
networks. In 1D lattices and Cayley trees it is trivial to follow a
minimum path because there is only one minimum path between two
arbitrary nodes. In 2D lattices, however, there are many minimum
paths. If a node can choose between two neighbors when sending a given
packet, and both neighbors belong to a minimum path between the origin
and the destination of the packet, one of them is chosen randomly with
equal probability. This algorithm is indeed the simplest one and the
interpretation of the results is clearer than for more complex routing
algorithms. However, congested nodes will still have lower probability
of receiving packets because of the definition of the quality of a
channel. Therefore packets will avoid those nodes to some extent, as
would happen in more sophisticated routing algorithms.

\section{Results}
For certain parameters of the model (in particular $\xi=1$
\cite{arenas01}) and for similar models of traffic \cite{ohira98}, a
transition from a {\it free} regime, where all the packets reach their
destination, to a {\it congested}, regime where some packets are
accumulated in the network, has been found. This transition is
properly described by means of the following order parameter $\eta(p)$
introduced previously \cite{arenas01}:
\begin{equation}
\eta(p)=\lim_{t\rightarrow\infty}\frac{1}{pS}\frac{\left\langle\Delta N\right\rangle}{\Delta t}
\label{def_order}
\end{equation}
In this equation $\Delta N=N(t+\Delta t)-N(t)$ and
$\langle\ldots\rangle$ indicates average over time windows of width
$\Delta t$. These averages can be over one or many realizations,
yielding the same result. Essentially, the order parameter represents
the ratio between undelivered and generated packets calculated at long
enough times such that $\Delta N\propto\Delta t$. Thus, $\eta$ is only
a function of the probability of packet generation per node and time
step, $p$.

The power spectrum $S(f)$ of the total number of packets in the
network $N(t)=\sum_i n_i$ is also used here to further understand the
phase transition. By means of the power spectrum the behavior of the
time correlations of the system can be studied.

Let us study separately the critical case $\xi=1$ and the non
critical cases $\xi<1$ and $\xi>1$.

\subsection{The critical case $\xi=1$}

A continuous transition between the free regime and the congested
regime occurs in hierarchical networks for $\xi=1$, as reported in
\cite{arenas01}. For small values of $p$, all the packets reach their
destination and the total number of packets $N(t)$ fluctuates around a
finite value. In this case the order parameter is $\eta=0$. However,
as $p$ increases, a critical point $p_c$ is reached, where the
fluctuations in $N(t)$ become very large and the characteristic time
of the system diverges (critical slowing down). Beyond this point,
some packets remain undelivered and $\langle N(t)\rangle $ grows
linearly with $t$. The same qualitative behavior is observed here for
1D and 2D lattices, although there are quantitative differences.

From an engineering point of view it is interesting to study first the
behavior of $p_c$ as the number of nodes in the network $S$ changes,
because it will provide valuable information about the scalability of
the network. Note that $N_c=p_cS$ is the maximum number of packets
that the network can handle per time step and, thus, it is a measure
of the capacity of the network.

For the hierarchical Cayley tree, a mean field calculation of $p_c$
was obtained in \cite{arenas01}: 
\begin{equation}
p_c^{CT}=\frac{\sqrt{z}}{\frac{z(z^{m-1}-1)^2}{z^m-1}+1}
\approx\frac{z^{3/2}}{(z-1)}\frac{1}{S}\mbox{ ,}
\label{pcB}
\end{equation}
where the approximation holds for $z^{m-1}\gg1$.

It is also possible to derive a mean field expression of $p_c$ for the
1D lattice. Since the most congested node is, from symmetry arguments,
the central one---the node at $\ell =L/2$---, the network will
collapse when the amount of packets received by this central node is
higher than the maximum number of packets that it is able to
deliver. Since in a large enough network it is safe to assume that the
central node will be congested similarly to its neighbors,
$n_{\ell-1}=n_{\ell}=n_{\ell+1}$, the maximum number of delivered
packets should be 1. On the other hand, the number of packets arriving
to the central node at each time step is the number of packets that
are generated at each time step at the left half of the network and
have their destination at the right half and conversely, this is
$pL/2$. Then the critical condition is given by
\begin{equation}
1=\frac{p_c^{1D}L}{2}\Rightarrow p_c^{1D}=\frac{2}{L}.
\label{pc1}
\end{equation}

\begin{figure}[t]
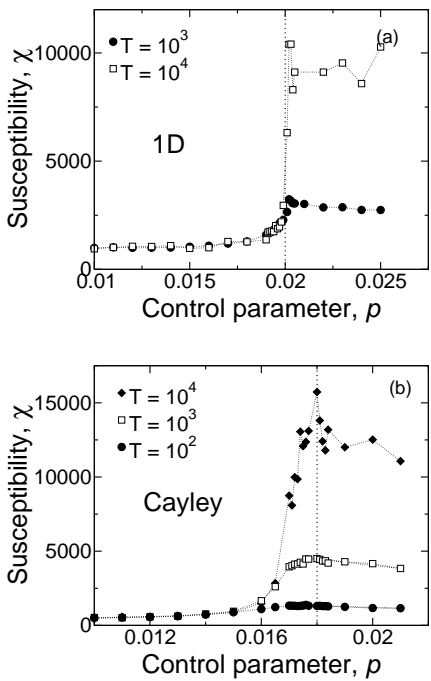

\includegraphics*[width=.65\columnwidth]{fig2a}\\
\vspace*{0.5cm}
\includegraphics*[width=.65\columnwidth]{fig2b}\\
\vspace*{-0.3cm}
\caption{Susceptibility for (a) a 1D lattice with $L=100$ and (b) a
(5,4) Cayley tree, for different time windows $T$. The vertical dotted
line corresponds to the mean field calculation of the critical point
(equations (\ref{pc1}) and (\ref{pcB}) respectively.)}
\label{suscep}
\end{figure}

These mean field expressions (\ref{pcB}) and (\ref{pc1}) can be
compared with simulations. Nevertheless, the fluctuations of $N(t)$
become very large near $p_c$ and it is difficult to calculate the
value of the order parameter. Instead, a susceptibility-like function
$\chi(p)$ can be defined by analogy with equilibrium thermal critical
phenomena, and used to estimate more accurately the value of the
critical probability of packet generation, $p_c$. The susceptibility
$\chi$ is related to the fluctuations of the order parameter by
\begin{equation}
\chi(p)=\lim_{T\rightarrow\infty}T\sigma_{\eta}(T)
\end{equation}
where $T$ is the width of a time window, and $\sigma_{\eta}(T)$ is the
standard deviation of the order parameter estimated from the analysis
of many different time windows of width $T$. Thus a calculation
implies a long realization of $N(t)$, its division into windows of
width $T$, calculation of the average value of the order parameter in
each window and finally the determination of the standard deviation of
these values. The susceptibility shows (Fig.~\ref{suscep}) a
singularity at $p_c$ as $T$ grows, as expected.

\begin{figure}[t]
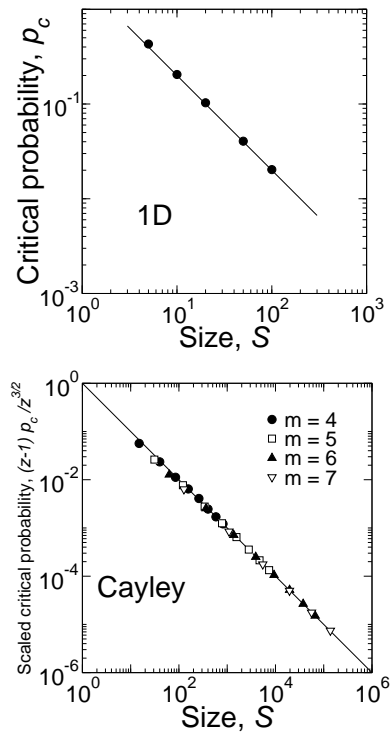

\includegraphics*[height=1.80in]{fig3a}\\ \vspace*{0.2cm}
\includegraphics*[height=1.90in]{fig3b}\\ \vspace*{-0.3cm}
\caption{Comparison between analytical (lines) and numerical (symbols)
values of $p_c$ obtained for the (a) 1D and (b) the Cayley trees. The
error bars of the numerical points are smaller than the size of the
symbols.}
\label{fig_pc}
\end{figure}

Fig.~\ref{fig_pc} shows that there is good agreement between
expressions (\ref{pcB}) and (\ref{pc1}) and the values of $p_c$
obtained numerically by means of susceptibility measures for different
network sizes.

\begin{figure}[t]
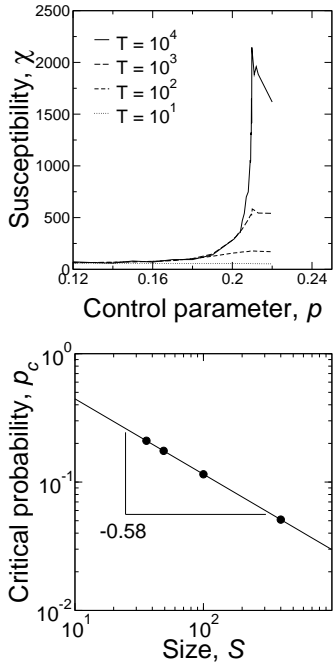

\includegraphics*[width=.5\columnwidth]{fig4a}\\
\vspace*{0.2cm}
\includegraphics*[width=.5\columnwidth]{fig4b}\\
\vspace*{-0.3cm}
\caption{(a) Susceptibility for a 2D $6\times 6$
lattice. (b) Scaling of the critical probability of packet generation
as a function of the size of the system for 2D square
lattices.}
\label{2D}
\end{figure}

For the 2D lattice it is more difficult to obtain even a mean field
expression for $p_c$. However, since for 1D lattices and Cayley trees
the scaling relation $p_c\propto S^{-1}$ holds---where $S$ is the size
of the network---, one may expect the same behavior for the 2D
lattice. Using the susceptibility to numerically determine $p_c$ from
simulations, one finds that this turns out to be incorrect. Although
it is difficult to obtain a precise value of the exponent,
Fig~\ref{2D} shows that it is close to 0.6 instead of 1.0:
\begin{equation}
p_c^{2D}\propto S^{-0.6}.
\end{equation}
This result suggests that the existence of multiple paths to get from
the origin to the destination has important consequences, not only in
shifting the value of $p_c$, but actually changing its critical
scaling behavior.

\begin{figure}[t]
\includegraphics*[width=0.9\columnwidth]{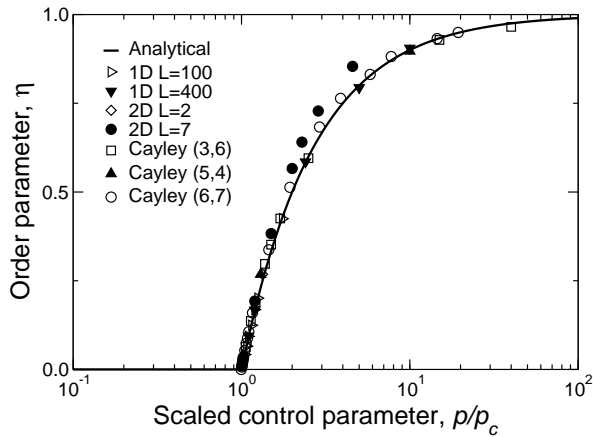}\\
\vspace*{-0.3cm}
\caption{Behavior of the order parameter in the case $\xi=1$. The
solid line corresponds to the analytical calculation for two nodes
exchanging information packets. Symbols correspond to simulations
performed in 1D, 2D and Cayley lattices.}
\label{par_ordre}
\end{figure}

The behavior of the order parameter is studied next. It is possible to
derive an analytical expression for the simplest 1D case where there
are only two nodes that exchange packets. Since from symmetry
considerations $n_1=n_2$, the average number of packets eliminated in
one time step is $2$, while the number of generated packets is
$2p$. Thus $p_c=1$ and with the present formulation of the model it is
not possible to reach the super-critical congested regime. However,
$p$ can be extended to be the average number of generated packets per
node at each step (instead of a probability) and in this case it can
actually be as large as needed. As a result, the order parameter for
the super-critical phase is $\eta=(p-1)/p$. As observed in
Fig.~\ref{par_ordre}, the general form
\begin{equation}
\eta(p/p_c)=\frac{p/p_c -1}{p/p_c}
\label{ord_eq}
\end{equation}
fits very accurately the behavior of the order parameter not only for
this simple 1D lattice with $L=2$ or the Cayley tree \cite{arenas01},
but also for any 1D lattice. Two-dimensional lattices again deviate
from this behavior, although the deviation is small.

In particular, near $p_c$, equation (\ref{ord_eq}) implies
\begin{equation}
\eta(p)\propto (p-p_c)
\end{equation}
and thus the critical exponent for the order parameter is equal to 1
at least for the 1D lattice and the Cayley tree.

\bigskip

\begin{figure}[t]
\includegraphics*[width=1.0\columnwidth]{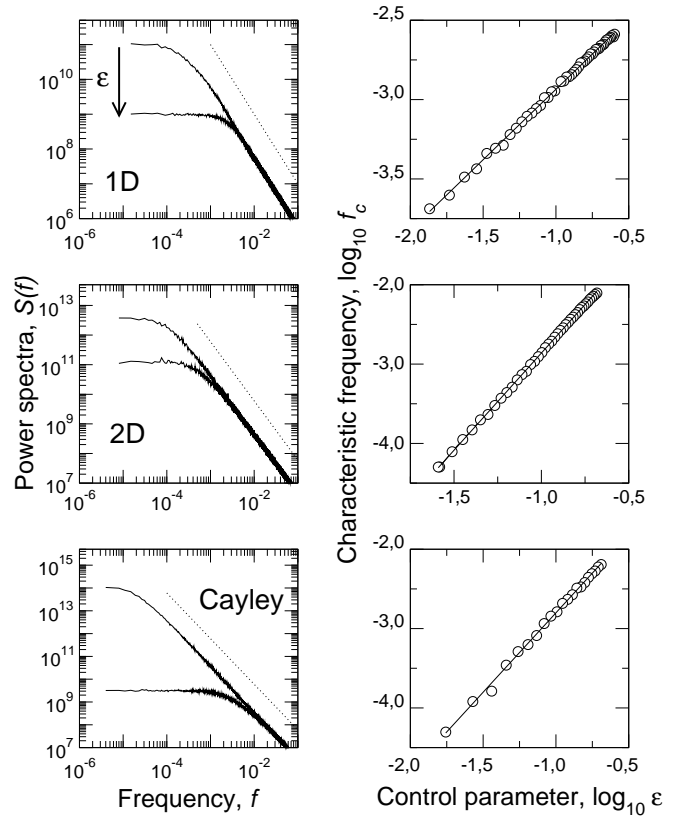}\\
\caption{Left: Log-Log plot of the power spectrum of $N(t)$ for
different topologies and different values of the control parameter
$\epsilon=(p_c-p)/p_c$ for the 1D case ($L=100$), 2D case ($L=6$), and
Cayley $(7,5)$. Power spectra have been obtained averaging over 100
realizations of $N(t)$. Dotted lines represent a power law with
exponent -2. Right: Characteristic frequency as a function of the
control parameter $\epsilon$ for the different topologies. As
observed, the characteristic frequency tends to 0 as $p\rightarrow
p_c$ following a power law. The straight lines correspond to fittings
of equation (\ref{forfc}).}
\label{spectra}
\end{figure}

The analysis of the power spectrum shows that in the sub-critical
regime, i.e. in the free phase, the spectrum is well fitted by a
Lorentzian characterized by a frequency, $f_c$, corresponding to
exponentially decaying correlations with a characteristic time $\tau$,
as depicted in Fig.~\ref{spectra}. As $p$ approaches $p_c$, it is
observed that $f_c\rightarrow0$ and the power spectrum becomes $1/f^2$
for the whole range of frequencies. Alternatively the characteristic
time diverges as $\tau\rightarrow\infty$ (critical slowing down). This
qualitative behavior is common to all the topologies of the underlying
network, as shown in Fig.~\ref{spectra}.

It is interesting to study how the characteristic frequency drops to 0
for each network topology. Near the critical point, one expects the
scaling behavior
\begin{equation}
f_c\propto (p_c-p)^\gamma.
\label{forfc}
\end{equation}
The value of the critical exponent $\gamma$ can be estimated by
fitting Eq.~(\ref{forfc}) to values of $f_c(p)$ close enough to the
critical point, as shown in Fig.~\ref{spectra}. Note that we fit
$p_c$ and $\gamma$ simultaneously. This procedure yields very
accurate values of $p_c$ but the values of $\gamma$ are subject to
large errors. Fig.~\ref{spectra} yields $\gamma\approx 0.9$ for a 1D
network with $L=100$, $\gamma\approx 2.5$ for a 2D network with $L=7$,
and $\gamma\approx 2$ for a (7,5) Cayley tree.

The determination of $\gamma$ is interesting not only from an academic
point of view, but also from an engineering perspective. Indeed, this
exponent is related to divergences of other relevant quantities near
the critical point. Any characteristic time $\tau$---the average time
to deliver a packet, for instance---will diverge as
\begin{equation}
\tau\propto(p_c-p)^{-\gamma}
\label{fort}
\end{equation}
and similarly the total number of packets
\begin{equation}
N\propto(p_c-p)^{-\gamma}
\end{equation}
since
\begin{equation}
\frac{N}{\tau}=pS
\end{equation}
from Little's law \cite{allen90}. This law states that, in steady
state, the number of delivered packets and the number of generated
packets are equal.

The estimation of $\gamma$ is particularly interesting in electronic
communication protocols. Indeed, Eq.~(\ref{fort}) is used to determine
the waiting time before a packet is considered lost in the network and
therefore sent again \cite{jacobson88}. In practice, the exponent
$\gamma=1$ predicted by classical queue theory \cite{allen90} is
assumed, while our current results suggest that more complex settings
can lead to exponents even larger than 2.

\subsection{Non critical cases $\xi<1$ and $\xi>1$}

\begin{figure}[t]
\vspace*{0.3cm}
\includegraphics*[width=0.75\columnwidth]{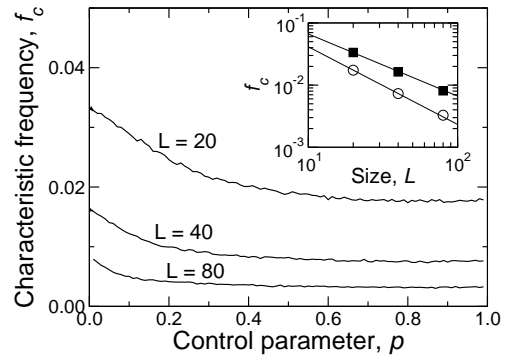}\\
\caption{Characteristic frequency $f_c$ as a function of the
probability of packet generation $p$, for $\xi=0.2$ and different
sizes of a 1D lattice. As observed, $f_c$ never becomes 0 as happens
in the critical $\xi=1$ case. Inset: Characteristic frequency at
$p\rightarrow 0$, $f_c^{0}$ (squares), and characteristic frequency at
large $p$, $f_c^{*}$ (circles). The lines represent the fittings
provided by equation (\ref{fc01d}) $f_c^{0}\propto L^{-1}$, and
equation (\ref{fc*1d}) $f_c^{*}\propto L^{-1/(1-\xi)}$, respectively.}
\label{fc_alt1}
\end{figure}

We have shown that the number of packets delivered by node $i$ is
$n_i^{1-\xi}$ and thus, when $\xi<1$, it increases with the number of
packets that this node accumulates. It is difficult to imagine a real
scenario with this characteristic. However, this case has been
included to understand the critical behavior when $\xi = 1$, i.e., to
show the relationship between criticality and the amount of packets
that can be delivered when load increases. As a consequence of the
increase of the deliver capability with the load, the transition to
collapse will never occur because, at some point in time, the number
of accumulated packets will be large enough and the number of
delivered and created packets will balance each other. Thus, the order
parameter will be zero for any value of the control parameter $p$, and
the correlations will decay exponentially. As shown in
Fig.~\ref{fc_alt1}, the characteristic frequency tends asymptotically
to $f_c^*$ as $p$ increases. This asymptotic value depends on the size
of the system.

For a 1D lattice with a high density of packets ($p\rightarrow 1$),
the number of packets that are delivered by a node is $n_i^{1-\xi}$
while the number of packets that are being delivered to this node is
proportional to $L$ (for instance, for the central node, this number
is simply $pL/2$). Therefore, $n_i\propto L^{1/(1-\xi)}$. The total
number of packets is $N=\sum_i n_i\sim L^{1+1/(1-\xi)}$ and according
to Little's Law
\begin{equation}
f_c^{*}\propto\frac{pL}{N}\propto L^\frac{-1}{1-\xi}
\label{fc*1d}
\end{equation}

On the other hand, for $p\rightarrow 0$ the scaling of the
characteristic frequency is given by
\begin{equation}
f_c^{0}\propto L^{-1}
\label{fc01d}
\end{equation}
since the packets success to jump from one node to the next at each
time step, and therefore the characteristic time is directly the
average path length between nodes.

Therefore, although there is no phase transition in this case $\xi<1$,
there is a cross-over from a low density behavior to a high density
behavior, as shown in Fig.~\ref{fc_alt1}. This crossover is also
observed in 2D lattices and Cayley trees.

\bigskip

The phase transition observed for $\xi=1$ is recovered when
$\xi>1$. Above a certain threshold, some packets are accumulated in
the network and the order parameter differs from 0. However, the
number of packets delivered by a node $i$ in this case $\xi>1$
decreases with the number of packets accumulated at that
node. Therefore, when some packets are accumulated, $n_i$ grows and
finally no packets are delivered at all. Thus suddenly above the
transition, which is discontinuous, the order parameter becomes 1.

\begin{figure}[t]
\vspace*{0.3cm}
\includegraphics*[width=0.45\columnwidth]{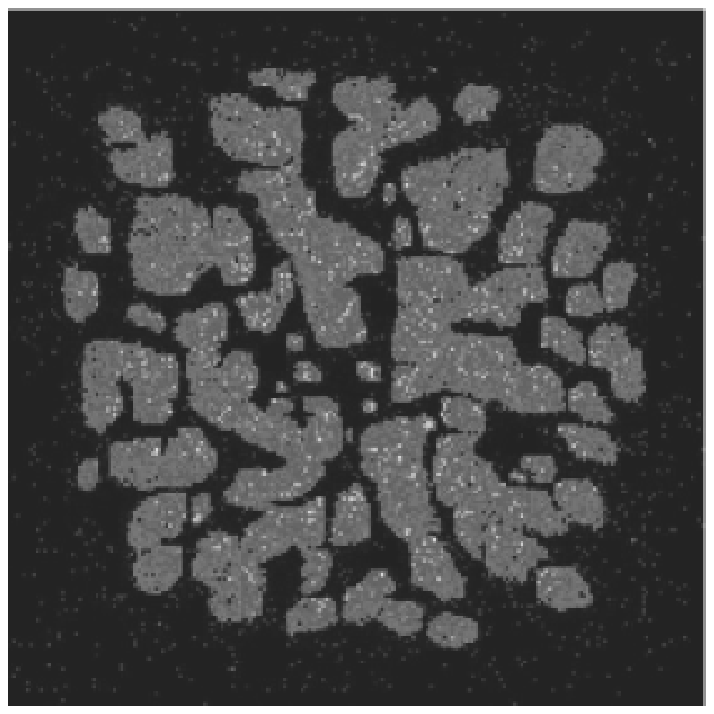}\quad\includegraphics*[width=0.45\columnwidth]{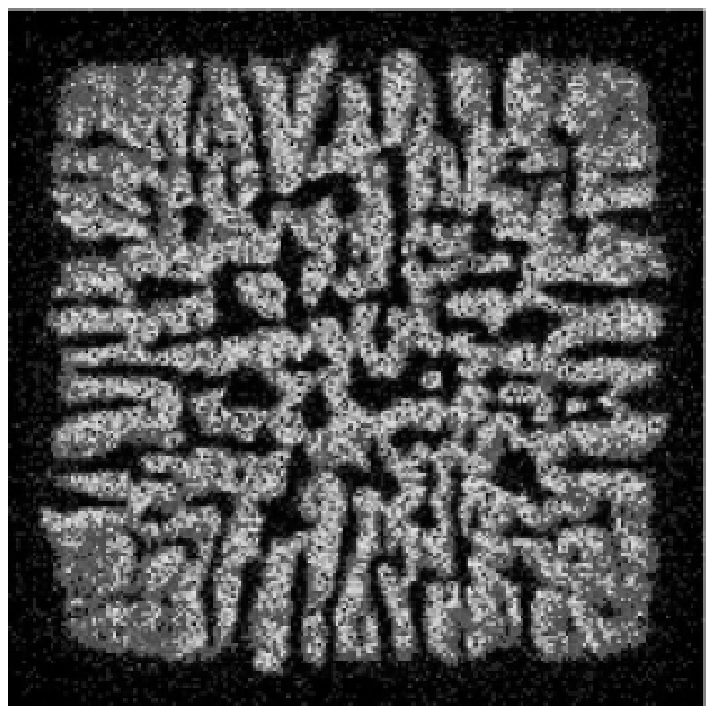}\\
\caption{Congestion nuclei formation for large 2D lattices of $L=200$,
in the non critical case $\xi>1$. Dark regions represent regions with
small congestion levels while bright regions correspond to highly
congested regions. Left: $\xi=5$ and $p=0.001$. Right: $\xi=2$ and
$p=0.01$.}
\label{patterns}
\end{figure}

The change in the order of the phase transition affects the spreading
of the collapse over the network. In the critical case $\xi=1$, the
collapse starts at the most {\it central} node and then spreads from
this point to the rest of the network. In this case $\xi>1$, the
reinforcement effect---the fact that the more collapsed a node is, the
more collapsed will get in the future---leads to the formation of many
congestion nuclei generated by fluctuations, that spread over the
whole network. Figure~\ref{patterns} illustrates the formation of
these congestion nuclei for 2D lattices with $\xi=5$ and $p=0.001$,
and $\xi=2$ and $p=0.01$

\section{Discussion and conclusions}
A collection of models recently proposed for hierarchical networks
\cite{arenas01} has been examined in detail for several network
topologies including 1D and 2D lattices, to characterize the phase
transition to collapse. New congestion scenarios have been analyzed.

First, it has been shown that the congestion behavior is governed by
the ability of agents to deliver information packets when their load
increases. When agents deliver packets at a constant ratio
independently of their load---number of packets to deliver---, a
continuous transition as the one reported in \cite{arenas01} for
Cayley trees is also observed for 1D and 2D lattices, the order
parameter being the fraction of accumulated packets. When the number
of delivered packets increases with the load, no phase transition is
observed. Conversely, when the number of delivered packets decreases
with the load, the order parameter jumps from zero to one and the
transition becomes discontinuous. These different behaviors are tuned
by a single parameter $\xi$, which is the exponent that determines how
the capability of nodes evolves with the number of accumulated
packets. When $\xi<1$ there is no transition to the congested regime,
for $\xi=1$ the transition is continuous and for $\xi>1$ the
transition is discontinuous. Note that the continuous transition
(reported in \cite{arenas01} and in some models of queues
\cite{tretyakov98,sole01}) is only a particular case between a
no-congestion behavior and a discontinuous transition behavior. Thus,
the critical behavior is intimately related to the independence
between load and deliver capability of nodes.

These different behaviors have been analyzed separately. The critical
case $\xi=1$ presents the most interesting behavior. To properly
understand the physics of the collapse process it is necessary to know
how the system approaches the critical point. The order parameter
becomes useless in this region due to the emergence of large
fluctuations, and then we have defined a susceptibility-like function
that is related, by analogy to equilibrium critical phenomena, to
fluctuations of the order parameter. This quantity shows a peak at the
critical point and allows to study the dependence on the size and
topology of the system. For 1D lattices and Cayley trees, it has been
shown (by means of simulation and also with mean field calculations)
that $p_c$ scales as the inverse of the size of the system,
$p_c\propto S^{-1}$. However, for 2D lattices a different scaling
$p_c\propto S^{-0.6}$ is obtained. It is suggested that the existence
of multiple paths from the origin to the destination of the packets is
the responsible for this change in the scaling behavior. On the other
hand, the study of the power spectrum of the number of the packets in
the network as a function of time has provided valuable information
about the variation of temporal correlations in the system. In
particular, below $p_c$ correlations decay exponentially, with a
characteristic time $\tau$ that diverges at $p_c$. The exponent of
this divergence is significantly larger than what is expected from
queue theory. This can be relevant to accurately forecast the temporal
behavior in complex communication settings.

For $\xi<1$, no phase transition is observed. Instead, a cross-over
from a low-density to a high-density regime occurs. In the low density
regime, the characteristic frequency is determined by the average
distance between nodes, while in the high density regime the
characteristic frequency is determined by the capability of agents to
deliver information packets. In 1D lattices, the cross-over changes
the scaling from $f_c\propto L^{-1}$ at low densities to $f_c\propto
L^{-1/(1-\xi)}$ at high densities.

Finally, when $\xi>1$ the transition to the congested regime is
discontinuous, since the number of delivered packets decreases with
the load and for long times no packets are delivered at all. Thus the
order parameter jumps from 0 to 1. Moreover, due to the reinforcement
effect---the fact that the more collapsed a node is, the more
collapsed it will get in the future---leads to the formation of many
congestion nuclei generated by fluctuations, that spread over the
whole network. The existence and distribution of such congestion
nuclei can also be of interest from an engineering point of view.

\acknowledgments The authors are gratefully acknowledged to
L.A.N. Amaral, A. Cabrales, L. Danon, X. Guardiola, C.J. P\'erez,
M. Sales, and F. Vega-Redondo. This work has been supported by DGES of
the Spanish Government, Grants No. PPQ2000-1339 and No. BFM2000-0626,
and EU TMR grants No. IST-2001-33493 and
No. ERBFMRXCT980183. R.G. also acknowledges financial support from the
Generalitat de Catalunya.


\begin{thebibliography}{10}

\bibitem{watts98}
D. Watts and S. Strogatz, Nature {\bf 393},  440  (1998).

\bibitem{barabasi99}
A.-L. Barabasi and R. Albert, Science {\bf 286},  509  (1999).

\bibitem{amaral00}
L.~A.~N. Amaral, A. Scala, M. Barthelemy, and H.~E. Stanley, Proc. Nat. Acad.
  Sci. {\bf 97},  11149  (2000).

\bibitem{albert02}
R. Albert and A.-L. Barabasi, Reviews of Modern Physics {\bf 74},  47  (2002).

\bibitem{dorogovtsev??}
S. Dorogovtsev and J.~F.~F. Mendes, Preprint cond-mat/0106144  (2001).

\bibitem{faloutsos99}
M. Faloutsos, P. Faloutsos, and C. Faloutsos, Comp. Comm. Rev. {\bf 29},  251
  (1999).

\bibitem{albert99}
R. Albert, H. Jeong, and A.-L. Barabasi, Nature {\bf 401},  130  (1999).

\bibitem{huberman99}
B. Huberman and L. Adamic, Nature {\bf 401},  131  (1999).

\bibitem{ebel??}
H. Ebel, L.-I. Mielsch, and S. Bornholdt, Preprint cond-mat/0201476  (2002).

\bibitem{adamic01}
L.~A. Adamic, R.~M. Lukose, A.~R. Puniyani, and B.~A. Huberman, Physical Review
  E {\bf 64},  046135  (2001).

\bibitem{radner93}
R. Radner, Econometrica {\bf 61},  1109  (1993).

\bibitem{decanio98}
S. DeCanio and W. Watkins, Journal of Economic Behavior and Organization {\bf
  36},  275  (1998).

\bibitem{jacobson88}
V. Jacobson,  in {\em Proceedings of SIGCOMM '88} (ACM, Standford, CA, 1988).

\bibitem{ohira98}
T. Ohira and R. Sawatari, Physical Review E {\bf 58},  193  (1998).

\bibitem{tretyakov98}
A. Tretyakov, H. Takayasu, and M. Takayasu, Physica A {\bf 253},  315  (1998).

\bibitem{arenas01}
A. Arenas, A. Diaz-Guilera, and R. Guimera, Physical Review Letters {\bf 86},
  3196  (2001).

\bibitem{sole01}
R. Sole and S. Valverde, Physica A {\bf 289},  595  (2001).

\bibitem{csabai94}
I. Csabai, Journal of Physics A: Mathematical and General {\bf 27},  L417
  (1994).

\bibitem{takayasu96}
M. Takayasu, H. Takayasu, and T. Sato, Physica A {\bf 233},  824  (1996).

\bibitem{guimera01}
R. Guimera, A. Arenas, and A. Diaz-Guilera, Physica A {\bf 299},  247  (2001).

\bibitem{allen90}
O. Allen, {\em Probability, statistics and queueing theory with computer
  science application}, 2nd ed. (Academic Press, New York, 1990).

\end{thebibliography}

\end{document}